\documentclass{article}

\begin{document}

\title{Consequence for Wavefunction Collapse Model of the Sudbury Neutrino Observatory Experiment}
\author{Gordon Jones, Philip Pearle and James Ring\\
Department of Physics, Hamilton College, Clinton, NY  13323\\
E-mail: gjones, ppearle, jring @hamilton.edu }

\maketitle

It is shown that data on the dissociation rate of deuterium obtained 
in an experiment at the Sudbury Neutrino Observatory provides evidence that the  
Continuous Spontaneous Localization wavefunction collapse model 
should have mass--proportional coupling to be viable.
 

\section{Introduction}\label{Section I}

	The ``measurement problem" can be thought of as indicating a serious flaw in 
the foundations of quantum theory\cite{Bell}, and therefore a potential clue to new physics. 
When a measurement is described by Schr\"odinger's equation, the statevector evolves to 
a superposition of apparatus states, each recording a different outcome. This conflicts with 
the interpretation of the statevector as describing ``reality" since only one 
of these apparatus states exists in reality. 
   
  One resolution of this problem is to modify Schr\"odinger's equation, adding a term so that 
the ``collapse" of the statevector to a macroscopically unique state occurs dynamically. 
The added term depends upon a randomly 
fluctuating quantity.  The particular realization of that quantity determines the particular 
macroscopic collapse outcome, and the statistics of the quantity gives the outcomes 
with the Born probabilities\cite{Pearle76,Pearle79}.  

	Arguably, the Continuous Spontaneous Localization (CSL) model\cite{PearleCSL,GPR+} 
is at present the most sucessful dynamical collapse model\cite{others}. In 
CSL, the fluctuating quantity is a scalar field.  In each added term it  
is coupled to the operator representing the number density of particles of a particular type  
(e.g., electrons, protons, neutrons...) by a coupling constant $g_{\alpha}$ 
(e.g., $g_{e}$, $g_{p}$, $g_{n}$...).  

The construction of 
CSL crucially depended upon the advent of a model of Ghirardi, Rimini and Weber (GRW)\cite{GRW}.    
In the GRW model, the overall rate of collapse for any isolated particle in a 
superposition of ``widely separated" states is chosen to be $\lambda=10^{-16}$ sec$^{-1}$.  
``Widely separated" refers to greater than the mesoscopic distance $a=10^{-5}$cm: for separation less than $a$, 
the collapse rate is less than $\lambda$.  These features of the GRW model 
are carried over into CSL, except that the rate of collapse 
for particles of type $\alpha$ becomes $\sim \lambda g_{\alpha}^{2}$. The GRW collapse 
rate $\lambda$ is assigned to the proton, so $g_{p}\equiv 1$: experimental limits on the parameters $g_{e}/g_{p}\equiv g_{e}$, 
$g_{n}/g_{p}\equiv g_{n}$ are discussed in this paper. 

	Since CSL is different than standard quantum theory, in some instances it makes different predictions 
which are experimentally testable. The most direct test is to put an object into a superposition of 
spatially separated states for a while, and then bring the states together and measure their 
interference\cite{Int} (see also the scheme of reference\cite{Henkel}). A sufficiently 
sensitive direct test has not yet been proposed.  
However, there are ``side effects" 
predicted by CSL which are testable.  One is that a small object will undergo random walk\cite{randomwalk}. Another  
is that a charged particle will be shaken and therefore radiate\cite{Fu}.  
Another, the one which concerns us here, is that a bound state will be ``spontaneously excited" 
by the collapse mechanism\cite{exc}. 

\vskip48pt

\section{Collapse--Induced Bound State Excitation}\label{Section II}

	Collapse narrows wavefunctions and so, by the uncertainty principle, the particles described by the 
narrowed wavefunction possess greater 
momentum and, therefore, energy. (The energy gain may be attributed to a loss of energy 
of the fluctuating field that causes the collapse\cite{Pearleenergy}.)  When particles are in a bound state, their  
wavefunction continually undergoes a slight modification.  This modified wavefunction may be written as 
the sum of a complete set of states, bound (with the coefficient of the original bound state 
of magnitude slightly less than one) and excited.  Those particles which are excited 
move apart, due to the ever-present usual Schr\"odinger evolution, and they can 
collide with and give their energy to other particles placed nearby for the purpose of detection.  
The probability/sec of excitation of a bound state $|\psi\rangle$, to an excited state $|\phi\rangle$ (normalized to 1), 
when expanded in a power series in (size of bound state/$a$), is\cite{PearleSquires}

\begin{equation}\label{2.1} 
{dP\over dt}={\lambda\over 2a^{2}}|\langle\phi|\sum_{\alpha,i}g_{\alpha}{\bf r}_{\alpha i}|\psi\rangle|^{2}
\thinspace +\thinspace 
\hbox{O}[\lambda {\rm (bound
\thinspace\thinspace 
state\thinspace\thinspace size}/a)^{4}], 
\end{equation}

\noindent where ${\bf r}_{\alpha i}$ is the position operator 
(or center of mass position operator for an extended particle like a nucleon) 
of the $i$th particle of type $\alpha$.   

	Eq. (\ref{2.1}) holds irrespective of the excitation of the center of mass of the bound state.  
If ${\bf Q}$ is the center of mass position operator, the states $|\psi\rangle$ and $|\phi\rangle$ 
are defined so that  

\begin{equation}\label{2.2} 
	{\bf Q}|\psi\rangle={\bf Q}|\phi\rangle=0,   
\end{equation}

\noindent where ${\bf Q}\equiv\sum_{\alpha,i} m_{\alpha,i}{\bf r}_{\alpha i}/\sum_{\alpha,i} m_{\alpha,i}$. 
An interesting consequence of Eqs. (\ref{2.1}), (\ref{2.2}) is that, for mass--proportional coupling 
($g_{\alpha}\sim m_{\alpha}$), the first term in (\ref{2.1}) vanishes identically\cite{PearleSquires}. 
Mass--proportional coupling is suggestive of a connection between collapse and gravity\cite{gravity}.  
There is an interpretation of CSL\cite{GhirWeb} which requires mass--proportional coupling, although 
other interpretations\cite{PearleAbner} are possible.  There also is an argument\cite{Alberto}  
applicable to a nonrelativistic collapse theory (i.e.,where mass and energy are distinctly different entities),
that the sum of couplings of the individual constituents of a compound object ought to be the coupling of the object, which  
may be implemented by mass-proportional coupling.  

	The term in (\ref{2.1}) $\sim$(size of bound state/$a$)$^{4}$ is too small to be observed in present experiments. 
However, the first term in (\ref{2.1}) is experimentally testable. A very low noise experiment exists which 
observes radiation appearing in a slug of Ge (order of a kg), over a period of time (order of a year)\cite{Avignone}. 
What is looked for here is collapse--induced ionization of a 1s electron in Ge, which should give 
a photon pulse of 11.1 keV (the binding energy of the 1s electron, emitted by the remaining electrons in the atom 
as they readjust) plus kinetic energy deposited 
in the Ge (which is also the detector) by the ejected electron.  Analysis of the experimental data gives\cite{CollettRing} 

\begin{equation}\label{2.3} 
	|g_{e}-{M_{e}\over M_{p}}|<12{M_{e}\over M_{p}}\bigg[{(\lambda/a^{2})_{\rm GRW}\over \lambda/a^{2}}\bigg]^{1/2}.
\end{equation}

\noindent So, if $\lambda/a^{2}$ has the GRW value, it follows from (\ref{2.3}) that $g_{e}$ is within 1200\% of 
mass proportionality:

\begin{equation}\label{2.4} 
	0\leq g_{e}<13{M_{e}\over M_{p}}.
\end{equation}

\vskip48pt

\section{Collapse--Induced Deuteron Excitation}\label{Section III}

	In this paper we consider observation of deuteron excitation in order to 
obtain a limit on $|g_{n}-M_{n}/M_{p}|$.  (In what follows we neglect the electron contribution.) 
For deuterons, Eq. (\ref{2.2}) implies that 
${\bf r}_{p}=-{\bf r}_{n}(M_{n}/M_{p})$ (more precisely, this operator equation 
holds when acting on $|\psi\rangle$ or $|\phi\rangle$), 
so the deuteron relative coordinate ${\bf r}\equiv {\bf r}_{p}-{\bf r}_{n}
=-{\bf r}_{n}[((M_{n}/M_{p})+1]$.  Then (\ref{2.1}) becomes 

\begin{eqnarray}\label{3.1}
{d\delta P\over dt}=&&
{\lambda\over 2a^{2}}|\langle\phi|{\bf r}_{p}+g_{n}{\bf r}_{n}|\psi\rangle|^{2}\\ \nonumber
=&&{\lambda\over 2a^{2}}\bigg[{g_{n}-M_{n}/M_{p}\over 1+M_{n}/M_{p}}\bigg]^{2}d{\bf k}
|\langle{\bf k}|{\bf r}|\psi\rangle|^{2}. 
\end{eqnarray} 

\noindent In Eq. (\ref{3.1}) we have replaced $|\phi\rangle\langle\phi|$ by $d{\bf k}|{\bf k}\rangle\langle{\bf k}|$ 
($|{\bf k}\rangle$ is the relative momentum eigenstate of the dissociated neutron and proton) and we have replaced 
$P$ in (\ref{2.1}) by $\delta P$ since there is an infinitesimal probability of excitation to any $|{\bf k}\rangle$ 
state. 

	In the experiment under consideration at the Sudbury Neutrino Observatory (SNO)\cite{sno}, 
deuterium (contained in heavy water inside a 12m diameter 
spherical shell) is dissociated by incident neutrinos via the neutral current reaction $\nu +d\rightarrow p+n+\nu $ 
(any flavor neutrino).  
The putative CSL deuteron dissociation occurs in addition to this. The free neutrons created by either mechanism  
are thermalized by travelling through the deuterium bath until they are either captured 
by a deuteron and form tritium (with the resulting emission of a 6.25 MeV gamma) or captured by 
$^{35}$Cl (since NaCl is dissolved in the heavy water) with typical emission of multiple gammas.      
The gammas Compton scatter electrons, whose Cerenkov radiation is detected by a surrounding array of 
photomultiplier tubes. Thus the experiment is not sensitive 
to the energy spectrum of the ejected neutrons, but only to their number.  Accordingly, we 
integrate Eq. (3.1) over all ${\bf k}$ to obtain 

 \begin{equation}\label{3.2}
{d P\over dt}={\lambda\over 2a^{2}}\bigg[{g_{n}-M_{n}/M_{p}\over 1+M_{n}/M_{p}}\bigg]^{2}
\langle \psi|r^{2}|\psi\rangle.  
\end{equation} 

	Various models\cite{nuc} give $\langle \psi|r^{2}|\psi\rangle\approx (3\cdot 10^{-13})^{2}$cm$^{2}$ to about 10\% 
accuracy.  If the volume of heavy water under observation is $V$ in units of thousand cubic meters and 
it is observed for $T$ years, we obtain from (\ref{3.2}) 
(taking the number of deuterons in heavy water to be $(2/3)\cdot10^{23}$/cc) the expected number of neutrons ejected from deuterium 
due to the CSL collapse mechanism:

 \begin{equation}\label{3.3}
N\approx 2.4\cdot 10^{7}{\lambda/a^{2}\over(\lambda/a^{2})_{GRW}}[g_{n}-M_{n}/M_{p}]^{2}TV.
  \end{equation} 
  
  \vskip48pt

\section{Comparison With Experiment}\label{Section IV}

	In the cited experiment,  
$T$=254.2/365=.70yrs and, since only events were accepted whose origin lay within a radius of 5.5m, then  
 $V=(4\pi/3)(5.5)^{3}/10^{3}=.70$km$^{3}$, so Eq. (\ref{3.3}) becomes

 \begin{equation}\label{4.1} 
 N\approx 1.2\cdot 10^{7}{\lambda/a^{2}\over(\lambda/a^{2})_{GRW}}[g_{n}-M_{n}/M_{p}]^{2}.
 \end{equation}

\noindent The number of detected neutron events was 1344.2 +69.8/-69.0 (statistical error) 
+98.1/-96.8 (systematic uncertainty)=1344.2 +120/-119. Here and below we are adding errors in quadrature.  
The reported detection efficiency is .40, which gives  
$N_{\rm expt}$=3361 +300/-298.  

	Since the Standard Solar Model (SSM) is well established, 
it is reasonable to consider the measured excess neutrons beyond those predicted by the  
SSM as due to CSL.  The SSM prediction of the $^{8}B$ flux \cite{Bahcall}, applied to the 
SNO experiment\cite{Ahmad}, is 13 +2.6/-2.08 neutrons/day or, for 254.2 days, $N_{SSM}$=3305 +661/-529 neutrons.  
Therefore, the excess number of neutrons which may be attributed to CSL is 

 \begin{equation}\label{4.2}
N_{CSL}\equiv N_{\rm expt}-N_{SSM}=56 +608/-725.
  \end{equation}
  
 \noindent That is, according to the experiment (to one sigma accuracy), the number 
 of excess neutrons produced due to the CSL excitation mechanism is  
 not larger than 664, although it could be as small as 0, in which case the theory is incorrect. 
 The negative value of the error in Eq. (9) simply quantifies the possibility that, in repeated 
 identical experiments, the mean value obtained by the experimenters could be less than was obtained in this 
 run. Of course, whatever mean value is measured, the putative CSL excitation would be a positive addition to the 
 neutrino excitation. Thus, it is the positive error 
 estimate that is relevant for our considerations.

	If we take from Eq. (\ref{4.2}) the experimental result $N_{CSL}<664$, we obtain from (\ref{4.1}) the inequality

 \begin{equation}\label{4.3} 
 |g_{n}-M_{n}/M_{p}|< 8\cdot 10^{-3}\bigg[{(\lambda/a^{2})_{GRW)}\over\lambda/a^{2}}\bigg]^{1/2}.
  \end{equation}
  
\noindent (The result is actually .0074 which we have rounded up to .008).  So, if $\lambda/a^{2}$ has the GRW value, 
it then follows 
from (\ref{4.3}) that $g_{n}$ is within 1\% of mass--proportionality: 

 \begin{equation}\label{4.4} 
 g_{n}={M_{n}\over M_{p}}\pm .008.
  \end{equation}
  
\noindent We note that a limiting factor in improving this result is 
the 20\% uncertainty in the SSM calculation which is the major source of the uncertainty in (11). Reduction 
of the combined theoretical and experimental uncertainty by a factor of about 30 is needed to 
reduce the uncertainty in $g_{n}$ shown in Eq. (11) to 
the order of $(M_{n}-M_{p})/M_{p}\approx 1.4\cdot 10^{-3}$\cite{Baryon}.  

	The result (11) for $g_{n}$ is 1600 times stronger than the comparable result (\ref{2.4}) for $g_{e}$.
 It strongly suggests that, if CSL with the GRW parameter values is to be physically viable, 
it should have mass--proportional coupling. 

	We conclude by considering other possible parameter values, since the GRW parameter values are ad hoc,   
although intelligently chosen.  In a recent paper, experimental and theoretical constraints on these parameters were 
discussed (see the last paper in reference \cite{randomwalk}).  
What is of interest here are the constraints on $\lambda/a^{2}$, where $(\lambda/a^{2})_{GRW}=10^{-6}$sec$^{-1}$cm$^{-2}$. 

	The upper limit $\lambda/a^{2}<2.5$sec$^{-1}$cm$^{-2}$ was obtained by Fu\cite{Fu} from comparison 
of the CSL prediction of radiation of electrons in the conduction band of Ge with experimental data\cite{Avignone}.

	The lower limit is a ``theoretical constraint."  The theory is supposed to 
describe the world as we see it, e.g., it should not leave uncollapsed for any ``appreciable amount of time" 
a superposed state of a ball in two ``discernibly different" places.  This constraint can 
be roughly quantified by requiring that a just visible object, a sphere of diameter $d\approx 4\cdot 10^{-5}$cm, 
in a superposition of two just touching states (centers separated by $d$), collapse 
in less than the human perception time $\approx 0.1$sec.The constraint formulas below are given in 
the first paper in reference\cite{CollettRing}.  

	For $a>>d/2$, the constraint is $[\lambda N^{2}d^{2}/4 a^{2}]^{-1}<.1$.    
$N$ is the number of nucleons in the sphere: for a sphere of density 1gm/cc, $N\approx 2\cdot 10^{10}$.  
This constraint becomes $\lambda/a^{2}>.6\cdot 10^{-10}$sec$^{-1}$cm$^{-2}$.   

	 For $a<<d/2$, the constraint is $[\lambda N^{2}a^{3}(4\pi)^{3/2}/V]^{-1}<.1$, where $V$ 
is the volume of the sphere.  This constraint becomes $\lambda/a^{2}>2\cdot 10^{-35}/a^{5}$sec$^{-1}$cm$^{-2}$.   
For $a\approx10^{-5}$cm (a value somewhat large for the applicability of this equation), this becomes 
$\lambda/a^{2}>2\cdot 10^{-10}$sec$^{-1}$cm$^{-2}$. 

	Therefore, for $(\lambda/a^{2})_{GRW}=10^{-6}$sec$^{-1}$cm$^{-2}<\lambda/a^{2}<2.5$sec$^{-1}$cm$^{-2}$, the error 
is even smaller than that in Eq. (11), making mass-proportional coupling even more likely.  On the other hand, at the 
``theoretical constraint" limit $\lambda/a^{2}\approx 10^{-10}$sec$^{-1}$cm$^{-2}$, the error in 
Eq. (11) becomes $\approx \pm .8$.  One would need an even better experimental constraint on the 
``spontaneous excitation" of bound states than given here to conclude in this case that   
mass-proportional coupling is inescapable for a viable collapse model.


\begin{thebibliography}{99}
\bibitem{Baryon} We note, until the uncertainty is reduced by that amount, the results are 
also consistent with Baryon number proportionality. This would imply e.g., that a macroscopically  
distinguishable superposition of leptons or of mesons would never collapse, and one might regard that 
as an undesirable feature of a collapse model.
\end{thebibliography}
\end{document}